\documentstyle{article}
\begin{document}
\begin{center}
{\Large \bf Single-particle subband structure of Quantum Cables}
\end{center}

\vspace{.3cm}
\noindent
\begin{center}
{\large Z. Y. Zeng$^{\ddagger}$, Y. Xiang, and L. D. Zhang}
\end{center}

\vspace{.001cm}
\noindent
\begin{flushleft}
{\it Institute of Solid State Physics, Chinese Academy of
Sciences, \rm P.O. Box 1129,  \it Hefei,\\ 230031, P. R. China \\}
\end{flushleft}

\vspace{.5cm}
\noindent
\begin{center}
{\bf Abstract}
\end{center}

We proposed a model of Quantum Cable in
analogy to the recently synthesized
coaxial nanocable structure [Suenaga et al. Science, 278, 653 (1997);
Zhang et al. ibid, 281, 973 (1998)], and studied its single-electron
subband structure. Our results show that the subband spectrum of
Quantum Cable is different from either  double-quantum-wire (DQW)
structure in two-dimensional electron gas (2DEG)
or single quantum cylinder.
Besides the double degeneracy of subbands
arisen from the non-abelian mirrow reflection symmetry,
interesting quasicrossings ( accidental degeneracies), anticrossings
and bundlings of Quantum Cable energy subbands
are observed for some structure parameters.
In the extreme limit (barrier width tends to infinity ), the normal
degeneracy of subbands
different from the DQW structure is independent on the other structure
parameters.

\vspace{1.cm}

\newpage

\noindent

\begin{center}
{\bf I. INTRODUCTION}
\end{center}

In recent years,  the energy spectrum and transport properties
of electrons in
low-dimensional systems have received extensive attention.
It is generally believed that quantum effects become more significant as
the system dimensionality is reduced.
In semiconductor, for example,
confining electrona in a $2$D plane, $1$D wire or
$0$D dot  gives rise to the obvious quantization of
electron motion, which eventually results in some unusual
transport and optical characteristics.
With these interesting properties people can realize nanoscaled
electronic devices with a variety of functions.

 Since the prediction that $1$D semiconductor quantum well wire
 can be of importance in high-speed-device applications[1], and their
 subsequent fabrication[2], there has been a great deal of interest in their
 transport  and optical properties.
 Constantinou et al. [3] investigated the single-electron
 energy subbands of a solid cyclindrical quantum wire in the absence and in
 the presence of an axial magnetic field. It is predicted [3]
 that the subband
 energy given by  finite confining potential
 is reduced compared with the values given by infinite
 confining potential in the absence of magneitc effect.
 When a magnetic field applied along the axis of the wire, a minimum in the
 energies associated with carriers have negative azimuthal quantum number.
 If a solid cylinder is replaced by a hollow cylinder, Masale et al. [4]
 found that,
 the application of  an axial magnetic field leads to a drastic modification
 in the subband spectrum.
 Makar et al.[5]
  demonstrated the oscillatory behavior of the density of states
 for a hollow cylinder under an axial magnetic field.
 Magneto-optical effects,
 collective excition, transport behavior and other physical features in
 cylindrical quantum wires
 were also actively studied [6].

    Coupled waveguide structures have long been the study subject of
 the optics  and microwave community.
  Compared with the single waveguide system, coupled
 waveguide structure possesses some striking and unique features arising from
 the coupling between two waveguides, such as the enhanced quantum confined
 Stark effect [7] etc. These unique
 features are very useful in producing numerous devices
 including digital switches, multipleexers, and
 tunable filters [8]. To exploit the analogy between electroamgnetic waves
 propagating along waveguides and electron transport in quantum wires,
 some theoretical investigations were devoted to the study of two coupled
 quantum
 wire used as field-effect directional couplers, energy filters, etc. [9].
 In 1990, two coupled
 quasi $1$D  quantum wires
 device was fabricated by Alamo and Eugster,
 and its transport properties were investigated both theoretically and
 experimentally. Later some groups
 calculated the ballistic conductance and magnetoconductance of
 such systems [11],
 and some interesting transport properties were displayed.
 Recently, Suenaga et al. [12] and Zhang et al. [13]
 synthesized successfully
 a new kind of quasi-one-dimensional composite
 structure termed as coaxial nanocalbe, in
 which two conducting cylinderical
 layers are seperated by a insulating layer.

    In the present work we propose a model of Quantum Cable  consisting of
    two  quantum cylinders  coupled through a controllable
    potential barrier, motivated partly by
    the recent successful synthesization of coaxial nanocable [12,13].
    Quantum Cable structure can be made either from the coaxial nanocable
    or from GaAs/AlAs system. For the former structure, if the
    energy-band structures of its two conducting layers are similar and
    the middle insulating layer is not thich enough to forbid electrons'
    tunneling, it can be viewed as a kind of Quantum Cable structure.
    For the later materials system, it is easier to fabricate and has
    close lattice matching. Depending on the Al cincentration in
    $Al_{1-x}Ga_xAs$, its band gap can be changed continuously, thus
    the shapes of the barrier and well can be made almost to what
    one desires [Guo , JAP].
    Therefore, Quantum Cable would be more easily fabricated
    from the GaAs/AlAs system. Moreover, multiple quantum cylindres
    and superlattice structure with cylindrical symmetry can also
    be available from the GaAs/AlAs systems.
    Quantum  Cable is similar to but different
    from the  DQW structure [10,11].
    The similarities
    lies in that they are both  coupled quantum-well structures. The
    major discrepance  between Quantum
    Cable and the DQW structure comes mainly from their
    different confining potential profiles and their symmetries.
     For DQW system,
    since extreme quantum limit is usually
    applied that only the lowest subband is
    populated, quantum confinement in one direction is enough to be
    taken into account.
    While for
    Quantum Cable system, its confining potential is symmetric with
    respect to Cable axis, and quantum confinement in two directions
    should be considered. This cylindrical symmetry introduces the
    two-fold degeneracies of eigen subbands, which consequently
    induces some unique optical and transport characteristics
    unexpected in the DQW system.
     Our numerical
    results show that the lowest (0,0) energy subband remains the ground
    subband either for single quantum
    cylinder ( solid and hollow) or Quantum Cable
    irrespective of their structure parameters. Here we used the
    azimuthal quantum number $n$ and radial quantum number $l$ to
    label the subband $(n,l)$ of Quantum Cable. As one of the structure
    parameters of Quantum Cable is varied with others being fixed,
    energy subbands of Quantum Cable exhibit interesting crossings
    (accidental degeneracies) and anticrossings (repulsions), in the
    meantime, energy difference between neighboring subbands display
    inhomogeneous variation. Therefore, Quantum Cable can be
    regarded as a concrete example for studying the diabolical points
    in energy-level surface of real Hamiltonian system [ ] and
    Berry phase accompanying adiabatic changes []. If one adjust the
    coupling between the two quantum cylindrical wires, one can
    observe subband bundling before the extreme limit arrives (
    i. e. the width of the coupling barrier tends to infinity).

   The paper is organized as follows. In Section II, we derived the
   formulas for calculating the energy subband in the single-electron
   approximation. Section III presents numerical calculations along with
   the associated analyses. A brief summary is given in Section IV.

\begin{center}
{\bf II. MODEL AND FORMULATION}
\end{center}

  The proposed Qunatum Cable comprises two coaxial cylindrical quantum
  wells.  They are coupled through a tunable potential barrier, which
  allows for electron's tunneling between two cylindrical wells.
   The confining
  potential of Quantum Cable is very similar
  to a recently fabricated structure of coaxial nanocable and schematically
  shown in Fig. $1$. The inside cylinder well has
  inner radius $R_1$ and outer radius $R_2$, the outside  cylinder has
  inner radius $R_3$ and outer radius $R_4$. The height and width of the
  coupling barrier are $U_B$ and $R_B$, respectively.
  It can be readily shown that the widths of the inside and outside
  cylinders are $R_{in}=R_2-R_1$ and $R_{ou}=R_4-R_3$.
  The electrons are free to move along the axis of Quantum Cable ,
   whereas their
  motion in the radial direction is quantized.
   In the effective
  electron mass approximation,  the Schr$\ddot{o}$dinger equation
  governing the motion of electron with energy $E$
  read as
  \begin{equation}
  [-\frac{\hbar^2}{2m^*} \bigtriangledown^2+U(\rho)]\Psi (\rho, \varphi,
  z)=E\Psi (\rho, \varphi, z),
  \end {equation}
where $m^* $ is the electron effective mass and
the confining potential of the Quantum Cable is
\begin{equation}
U(\rho) = \left\{ \begin{array}{ll}
               \infty, &\mbox {$\rho \leq R_1 \hspace{0.12cm} or \hspace{0.12cm}\rho \geq R_4$},  \\
                  V_0, &\mbox{$R_2 \leq \rho \leq R_3$}, \\
                    0, &\mbox{$otherwise$}, \end {array} \right.
\end {equation}
In the cyclindrical coordinates $(\rho,\varphi, z)$, the wave function
 $\Psi(\rho, \varphi, z)$ has the
  form  $\chi(\rho)e^{in\varphi} e^{ik_z z},
n=0,\pm 1, \pm 2,\cdots$,
  where $k_z$ is the axial wavevector.
 The raidal wave function $\chi(\rho)$
 satisfies the following Bessel equation
 \begin{equation}
 \rho^2 \frac{d^2 \chi}{d\rho^2} +\rho \frac{d\chi}{d\rho}
 +\{[2m^*(E-U(\rho))/\hbar^2-k_z^2]\rho^2-n^2\}\chi=0,
 \end{equation}
which has the following solutions
\begin{equation}
\chi(\rho)=\left \{ \begin{array} {llll}
A_nJ_n(k_1\rho)+B_nY_n(k_1\rho), & \mbox{$R_1\leq \rho \leq R_2$},\\
C_nK_n(k_2\rho)+D_nI_n(k_2\rho), & \mbox{$R_2 \leq \rho \leq R_3$},\\
F_nJ_n(k_1\rho)+G_nY_n(k_1\rho), & \mbox{$R_3\leq \rho \leq R_4$},\\
0,                               & \mbox{$ \rho \leq
R_1\hspace{0.12cm} or \hspace{0.12cm} \rho \geq R_4$},
\end{array} \right.
\end {equation}
where $J_n$ is the Bessel function of first kind, $Y_n$ the Bessel function
of second kind and $K_n, I_n$ are the modified Bessel functions, respectively,
and
\begin{eqnarray}
k_1 & = & [(2m^*_1/\hbar^2)E-k^2_z]^{1/2}, \nonumber \\
k_2 & = & [(2m^*_2/\hbar^2)(U_B-E)+k^2_z]^{1/2},
\end {eqnarray}
are the wavevectors with $m^*_i(i=1,2)$  being the electron effective mass
in medium $i$.  We now
apply the standard effective-mass boundary conditions at $\rho=R_1, R_2,
R_3,R_4$, which lead to the following equation satisfied by the eigen
energy subbands:
\begin{eqnarray}
 \frac{k_2}{m^*_2}{\it f_1}(k_1;R_1,R_2)[\frac{k_2}{m^*_2}
{\it F_1}(k_1,k_2;R_2,R_3,R_4)+\frac{k_1}{m^*_1}
{\it F_2}(k_1,k_2;R_2,R_3,R_4)]+  \nonumber \\
\frac{k_1}{m^*_1}{\it g_1}(k_1;R_1,R_2)[\frac{k_2}{m^*_2} {\it
G_1}(k_1,k_2;R_2,R_3,R_4)+ \frac{k_1}{m^*_1} {\it
G_2}(k_1,k_2;R_2,R_3,R_4)]=0,
\end{eqnarray}
where
\begin{eqnarray}
 {\it f_1}(k_1;R_1,R_2)   & = &  J_n(k_1 R_2)Y_n(k_1 R_1)-J_n(k_1 R_1)
                                     Y_n(k_1 R_2),  \nonumber \\
 {\it g_1}(k_1;R_1,R_2)   & = &  J_n(k_1 R_1)Y'_n(k_1 R_2)-J'_n(k_1 R_2)
                                     Y_n(k_1 R_1),  \nonumber \\
{\it F_1} (k_1,k_2;R_2,R_3,R_4) & = &  [K'_n(k_2 R_3)I'_n(k_2
R_2)-K'_n(k_2 R_2)
                                    I_n(k_2 R_3)]  \nonumber \\
                          & \times &  [J_n(k_1 R_4)Y_n(k_1 R_3)-J_n(k_1 R_3)
                                     Y_n(k_1 R_4)],  \nonumber \\
{\it F_2}(k_1,k_2;R_2,R_3,R_4) & = & [K_n(k_2 R_3)I'_n(k_2
R_2)-K'_n(k_2 R_2)
                                I_n(k_2 R_3)] \nonumber \\
                          & \times & [J'_n(k_1 R_3)Y_n(k_1 R_4)-J_n(k_1 R_4)
                                      Y'_n(k_1 R_3)], \nonumber \\
{\it G_1}(k_1,k_2;R_2,R_3,R_4) & = &  [K'_n(k_2 R_3)I_n(k_2
R_2)-K_n(k_2 R_2)
                                 I'_n(k_2R_3)] \nonumber \\
                          & \times &  [J_n(k_1 R_4)Y_n(k_1 R_3)-J_n(k_1 R_3)
                                  Y_n(k_1 R_4)], \nonumber \\
{\it G_2}(k_1,k_2;R_2,R_3,R_4) & = &    [K_n(k_2 R_3)I_n(k_2
R_2)-K_n(k_2 R_2)
                                    I'_n(k_2 R_3)] \nonumber \\
                          &\times &  [J'_n(k_1 R_3)Y_n(k_1 R_4)-J_n(k_1 R_4)
                                   Y'_n(k_1 R_3)],
 \end{eqnarray}
 where $f'(x)=df(x)/dx$.
 Equation $(6)$ may be solved numerically by employing the recursion
 relations satisfied by the Bessel functions outlined in Appendix.
It should be pointed out that the formulas presented above are exact since
no other approximation is made than the single-electron approximation.

\begin{center}
{\bf III. RESULTS AND DISCUSSIONS}
\end{center}

  To start with, we study the subband spetrum of single quantum
  cylinder.
  For the convinience
  of comparison, throughout this work,
   we take $m^*_1=5.73\times 10^{-32} kg, m^*_2=1.4 m^*_1$
  $m^*_3=m^*_1$, and
   energy in unit of $0.19 eV$ and length
  in unit of angstrom. Fig. $2$ shows the relation of subband energy with
  the solid quantum cylinder radius $R_3$ ( $R_1=R_2=R_3$).
  The subband energy $E_{nl}$ is related to the total energy
  via $E=E_{nl}+\hbar^2k^2_z/(2m^*)$ and is labeled by the azimuthal quantum
  number $n$ and the radial quantum number $l$. For simplicity,
  calculations are performed by setting $k_z=0$.
  As expected, the subband energy as well as the energy spacing between
  neighboring subbands decreases with the
  increase of the cylinder radius $R_3$; energy difference between
  adjoining subbands of lower-order
  is greater than that of higher-order.
   One can  see from Fig. $2$ that the lowest
  subband corresponds to $n=0$,
  the next bound subband is an $n=1$ subband.
  It is clear that for a given $n$, the value of $R_3$
  at which a confined subband
  appears satisfies $J_n(k_1R_3)=0$.
   It so happens that the first $n=0$ subband
  appears at a radius of about $40 \AA~$.  In Fig. $3$ we present
  the calculated subband energy of a hollow quantum cylinder
  against the outside radius $R_3$ with
   a given inside radius $R_2=100\AA~$,
  while the subband energy as a function of $R_2$ for a fixed $R_3=200\AA~$
 is given in Fig. $4$. In two cases, $R_1=0$ and $V_0=\infty$.
  It is obvious
  that the  subbands with equal radial quantum number $l$  converge
  as $R_3-R_2$ approaches zero, with their energies having an
  $(R_3-R_2)^{-2}$
  variation. As the inside radius  $R_2$ tends to zero while the outside
  radius $R_3$ keeps unchanged, we can observe an appreciable seperation
   between subbands corresponding to different $n$ but equal $l$
  quantum numbers. Crossings ( accidental degeneracies )
  of some excited subbands is also seen for fixed $R_3$ and
  varied $R_2$. However,
  the subband $(0,0)$ remains the ground subband in either of the two
  cases.
  In addition, the subbands with nonzero azimuthal quantum number
  are doubly degenerate ($E_{nl}=E_{-nl}$) for both solid cylinder and
  hollow cylinder structures, the application of a magnetic field
  will violate such degeneration [4]. The double degeneracy of
  subbands is due to the fact that the mirror reflection across
  the plane parellel to Cable axis is non-abelian [14].
  According to the above analysis, our results obtained from
  Eqn. $(6)$ is in very good agreement with Constantinou group's [3,4],
   which
  demonstrated the reliability of our formulation.

   Now we inspect the energy subband structure of Quantum Cable.
   The variation of subband energy with the outer
   cylinder radius $R_3-R_2$ is given in fig. $5$
   and with inner cylinder
   radius $R_1$ in Fig. $6$.
   The parameters are chosen such that $V_0=0.19 eV$,
   the barrier width $R_b=25\AA~$ and the inner cylinder radius
   $R_1=50 \AA~$ in Fig. $5$ and the outer cylinder radius $R_3-R_2=50\AA~$
   in Fig. $6$.  The subband (0,0)
   remains the ground subband and (1,0) the first excited subband whatever
the value of the outer or inner cylinder radius.
   In Fig. $5$, we also find,  as $R_3-R_2$ increases, the energy of
  some subband  drops very slowly while that of the other
   subbands falls relatively rapidly. This leads to the crossings
    ( accidental
   degeneracies ) of some
    energy subbands:  the second excited subband $(0,1)$
   as $R_3-R_2<60\AA~$  becomes the  third excited subband with its crossed
   subband (2,0) becoming the second excited subband when
   $R_3-R_2$  exceeds $60\AA~$ ;
   as the outer cylinder radius $R_3-R_2$ is further increased,
   it  will turns into the other higher-order
   excited subband while the
   crosssed subband is changed into the lower-order excited
   subband.   Certainly, other higher-order subbands also exhibit
   such accidental degeneracy phenomenon as can be found in Fig. $5$.
   In fig. $6$, we find  more interesting
   complicated accidental degeneracies:
   some excited subbands such as (0,1) and
   (1,1) becomes the lower-order excited subband first and then revives to
   the higher-order subband as the inner cylinder radius further
   increases.
   Another important feature is the inhomogeneous variation of
   energy intervals between adjoining subbands, which could be expected
   to result in
   some interesting optical and transport phenomena. It can be also expected
   that, the energy difference between neighboring subbands will be
   decrease with the further increase of either inner or outer cylinder
   radius.

   Due to the quantum tunnel effect,  electron's wavefunction
   will not be located within only one of the quantum cylinders,
   it would extend to the
   whole region of Quantum Cable.
   If the barrier height $V_0\rightarrow \infty$ or
   barrier width $R_b \rightarrow \infty$,
   Quantum Cable will become the
   simple structure of two seperated quantum cylinders.
   As the barrier height $V_0\rightarrow 0$ or
   barrier width $R_b \rightarrow 0$, Quantum Cable
   turns into  single solid quantum cylinder.
   It is therefore interesting
   to investigate how the coupling barrier
   influences the subband structure of Quantum
   Cable. In Fig. $7$
   we plot subband energy vs the barrier
   height $V_0$ and vs barrier width
   $R_b$ in Fig. $8$. With the increase of $V_0$, subband energy
   grows up, energy spacing between neighboring subbands varies
   inhomogeneously, accidental degeneracy of
   subbands is seen. As $V_0$ is further increased, we can
   observe accidental degeneracy of more than three subbands.
   While in the case of increasing barrier
   width, energies of some lowest subbands is lifted up while
   some falls off as $R_b$ increase.
   This reflects the fact that the coupling of two cylinder
   wells becomes weak for large $R_b$.
   With the further increment of the
   barrier width, subbands energy tends to keep constant and
   some subbands tends to degenerate into one subband, while the
   ground subband does not degenerate with other subband. It is
   heuristic to compare this feature with that of $2$D DQW structure.
   If the $2$D DQW system consists of two identical wires, subband
   degeneracy only happens to two subbands with the same symmetry
   in the extreme limit ($R_b\rightarrow \infty$),
   and the ground subband will also degenerate with
   one of the subbands [15]. For $2$D DQW with
   asymmetric confining potential, since the wave functions have
   no particular symmetry, no subband may cross and be degenerate.
   Though our calculating results are for Quantum Cabel with
   two cylinders of the same radius, it is expected that
   subband crossing and degeneracy also occurs in Quantum Cable of
   different radius cylinders, because the same symmetry
   still preserve in the structure.

   Finally,  we give the number of
   available subbands for Fermi energy below
   $0.19$ eV of a function of barrier height in Fig. (9a) and of barrier
   width in Fig. (9b). As expected, the available subband number exhibits
   an decreasing stepwise structure with the increase of barrier height
   and increasing stepwise profile as the barrier width increases.
   In addition, some comparatively narrow plateaus are also observed.
   This reflects the
   inhomogeneous variation of energy interval between adjoining subbands
   as the coupling paramater is varying.

\begin{center}
{\bf IV. CONCLUSIONS}
\end{center}

We studied the subband spectrum of Quantum Cable consists of a solid
quantum cylinder and a hollow quantum cylinder, which are coupled through
a potential barrier. The  subband energy of solid quantum cylinder,
hollow quantum cylinder and Quantum Cable are calculated.
The results demonstrated a fact that Quantum Cable is an unique system
to study quantum effects as bundling, accidental and normal degeneracy of
levels. For single quantum cylinder or Quantum Cable,
the subband $(0,0)$ always keeps the ground
subband whatever their structure parameters.
Accidental degeneracies of subbands could be displayed
 in the case of hollow cylinder
and Quantum Cable but does not appear in the subband spectrum of single
solid quantum cylinder. This phenomenon might be explained according
to Wigner-Von Neumann theorem [16]:
accidental degeneracy occurs only for the hermitian hamiltonian of no
less than three changable parameters. As one of the parameters of Quantum
Cable is varied, its energy subbands exhibit some interesting phenomena
such as bundling, accidental and normal degeneracy, and
inhomogeneous variation of energy seperation between adjoining subbands.
These features can be expected to be revealed in the optical
spectrum observation of Quantum Cable.
It is noted that accidental and normal degeneracy always
occurs whatever the value of the
inner-cylinder radius and outer-cylinder radius.
It is contrast to the case of $2$D DQW structure
in which subband degeneracy occurs only for $2$D DQW system
with symmetric confining potential, since the wavefunctions
have no particular symmetry  in the asymmetric potential case.
While for Quantum Cable structure, whatever the inner-wire radius and
outer-wire radius, the symmetry is always preserved.
Comparatively narrow plateaus are also observed in the plots of
the available subband number as a function of the barrier height or
width for a given Fermi energy, which will be reflected in the
transport properties of electrons along the Cable axis. Subband
spectrum of Quantum Cable with the application of magnetic field and
other properties deserve further investigations.

\begin{center}
{\bf ACKNOLEDGEMENT}
\end{center}

  This work is supported by a key project for fundamental research
in the National Climbing Program of China.

\begin{center}
{\bf APPENDIX}
\end{center}

In this appendix, we give some relations satisfied by various
kinds of Bessel functions:
\begin{eqnarray*}
\begin{array}{cc}
J_{-n}(x)=(-1)^nJ_n(x), &  Y_{-n}(x)  =  (-1)^nY_n(x),   \\
I_{-n}(x)=I_n(x) ,   &  K_{-n}(x)  =  K_n(x) ,        \\
J'_n(x)=J_{n-1}(x)-nJ_n(x)/x, & Y'_n(x)    =  Y_{n-1}(x)-nY_n(x)/x,  \\
I'_n(x)=I_{n-1}(x)-nI_n(x)/x, & K'_n(x)    =  -K_{n-1}(x)-nK_n(x)/x \\
\end {array}
\end{eqnarray*}
where $n$ is the natural number.
The above equations are useful in evaluating the Bessel functions.

\vspace{.5cm}
\noindent
{\bf References}

\vspace{.1cm}
\noindent
\hspace{0.2cm} $\ddagger$ E-mail address: zyzeng@mail.issp.ac.cn
\begin{enumerate}
\item Sakaki H 1980 Japan. J. Appl. Phys. {\bf 19} L735.
\item Petroff P M, Gossard A C, Logan R A and Wiegmann W 1982
Appl. Phys. Lett. {\bf 41} 635;
 Thornton T J, Pepper M, Ahmed H, Andrews D and Davies G J 1986
Phys. Rev. Lett. {\bf 56} 1198; Cibert J, petroff P M, Dplan G J,
 Pearton S J, Gossard A C and English J H 1986 Appl. Phys. Lett. {\bf 49}
 1275; Temkin H, DoalnG J, Parish M B and Chu S N G 1987
  ibid. {\bf 50} 413.
\item Constantinou N C and Ridley B K 1989  J. Phys.: condens. Matter.
{\bf 1} 2283; Constantinou N C, Massale M and Tilley D R 1992
ibid. {\bf 4} 4499.
\item
 Massale M, Constantinou N C and Tilley D R 1992
Phys. Rev. B {\bf 46} 15432.
\item  Makar M N, Ahmed M A and Awad M S 1991
Phys. Stat. Sol. (b) {\bf 167} 647.
\item  Chen H, Zhu Y and Zhou S 1987 Phys. Rev. B {\bf 36} 8189;
Huang F Y 1990 Phys. Rev. B {\bf 41} 12957;
Wendler L and Grigoryan V G 1994 Pgys. Rev. B {\bf 49} 14531;
Pokatilov E P, Klimin S N, Balaban S N and Fomin V M 1995
Phys. Stat. Sol. (b) {\bf 189} 433.
\item Kroemer H and Okamoto 1984 Jpn. J. Appl. Phys. {\bf 23} 970.
\item  Yariv A 1985  Optical Electronics ( HRW, New York ).
\item del Alamo J A and Eugster C C 1990 Appl. Phys. Lett. {\bf 56} 76;
Yang R Q and Xu J M 1991 Phys. Rev. B {\bf 43} 1699; Zhao P 1992 Phys.
Rev. B {\bf 45} 4301; Xu G, Yang M and Jiang P 1993 J. Appl. Phys.
{\bf 74} 6747.
\item Eugster C C, del Alamo J A, Rooks M J and Melloch M R 1992
Appl. Phys. Lett. {\bf 60} 642;
Eugster C C, del Alamo J A, Melloch M R and rooks M J 1992
Phys. Rev. B {\bf 46} 10406;
Eugster C C, del Alamo J A, Melloch M R and rooks M J 1993
ibid {\bf 48} 15057.
\item Wang J, Guo H and R. harris 1991 Appl. Phys. Lett. {\bf 59} 3075;
Wang J, Wang Y J and R. harris 1992 Phys. Rev. B {\bf 46} 2420;
Shi J R and Gu B Y 1997 Phys. Rev. B {\bf 55} 9941.
\item Suenaga K, Colliex C, Demoncy N, Loiseau A, Pascard H and
Willaime F 1997 Science {\bf 278} 653.
\item Zhang Y, Suenaga K, Colliex C and Iijima S  1998 Science {\bf 281}
 973.
\item  Zeng J Y 1997 Quantum Mechanics ( in Chinese ) (Scientific Press).
\item Kamizato T and Matsuura M 1989 Phys. Rev. B {\bf 40} 8378;
Morrison M A, Estle T L and Lane N F 1976 Quantum States of Atoms,
Molecules, and Solids (Prentice-Hall Inc.).
\item wigner E P and von Neumann J 1929 J. Phys. Zeif {\bf 30} 467.

\end{enumerate}

\newpage
\noindent
\begin{center}
{\bf Figure Captions}
\end{center}
\vspace{0.5cm}
\noindent
{\bf Fig.~1}~ Schematic view of Quantum Cable structure.

\vspace{0.5cm}
\noindent
{\bf Fig.~2}~   Lowest-oder energy subbands of solid quantum cylinder
     as a function of radius.

\vspace{0.5cm}
\noindent
{\bf Fig.~3}~   Lowest-oder energy subbands of hollow quantum cylinder
     as a function of outside radius.

\vspace{0.5cm}
\noindent
{\bf Fig.~4}~   Lowest-oder energy subbands of hollow quantum cylinder
     as a function of inside radius.

\vspace{0.5cm}
\noindent
{\bf Fig.~5}~   Lowest-oder energy subbands of Quantum Cable
     as a function of outer-wire radius $R_3-R_2$.

\vspace{0.5cm}
\noindent
{\bf Fig.~6}~   Lowest-oder energy subbands of Quantum Cable
     as a function of inner-wire radius $R_1$.

\vspace{0.5cm}
\noindent
{\bf Fig.~7}~   Lowest-oder energy subbands of Quantum Cable
     as a function of the coupling barrier height $V_0$.

\vspace{0.5cm}
\noindent
{\bf Fig.~8}~   Lowest-oder energy subbands of  Quantum Cable
     as a function of coupling barrier width $R_b$.

\vspace{0.5cm}
\noindent
{\bf Fig.~9}~   Number of available subbands for Fermi energy below $0.19$ eV
as a function of (a) coupling barrier height $V_0$ and (b) coupling
barrier width $R_b$.

\end {document}